# Role of Particle Shape in Strain-Controlled Resuspension of Dense Suspensions

Mohammadreza Mahmoudian,[1] Mahdi Vaezi,[2] and Parisa Mirbod*[1]
[1)] Department of Mechanical and Industrial Engineering, University of Illinois Chicago, Chicago, IL 60607, USA
[2)] Department of Engineering Technology, Northern Illinois University, DeKalb, IL, USA
(*Electronic mail: pmirbod@uic.edu.)

Dense non-Brownian suspensions exhibit complex resuspension dynamics governed by hydrodynamic interactions, particle microstructure, and gravitational confinement. While recent studies have shown that viscous resuspension in spherical suspensions is controlled by accumulated strain, whether this strain-based framework extends to anisotropic particles remains unresolved. Here, we investigate the role of particle shape in shear-induced resuspension using dense suspensions of spherical and rod-shaped particles subjected to steady and oscillatory shear. The results reveal that viscous resuspension consists of two distinct strain-controlled transitions: particle detachment from the sediment bed and transition to a fully suspended state. Across a broad range of normalized volume fractions, both particle shapes exhibit a nearly identical critical strain for detachment, $\gamma_{f,c1} \approx 6$, suggesting that resuspension onset is governed primarily by a local mobilization criterion that is relatively insensitive to particle morphology. In contrast, the strain required to reach a fully suspended state depends strongly on particle shape. Spherical suspensions reach a fully suspended state at $\gamma_{f,c2} \approx 120$, whereas rod suspensions require approximately 50% greater accumulated strain ($\gamma_{f,c2} \approx 180$). A fluid-strain scaling collapses the onset of resuspension across particle concentrations and particle shapes while highlighting the strong shape dependence of the transition to a fully suspended state. These findings establish a unified strain-based framework for viscous resuspension and demonstrate that particle anisotropy influences not the initiation of particle mobilization, but the subsequent pathway toward complete resuspension.

## I. INTRODUCTION

Shear-induced migration and resuspension of particles in dense non-Brownian suspensions play a central role in a wide range of natural and industrial processes, including sediment transport, slurry handling, and particle-laden flows in confined geometries. Under imposed shear, suspended particles undergo collective rearrangements driven by hydrodynamic interactions, particle contacts, and gradients in stress and concentration, leading to migration phenomena in canonical configurations such as Couette, cone-and-plate, and parallel-plate flows[1–5]. In negatively buoyant suspensions, where particle density exceeds that of the suspending fluid and gravity promotes sedimentation, shear can re-entrain particles from a sediment bed through a process known as *viscous resuspension*[6–8]. This process occurs when viscous stresses generated by the flow overcome gravitational forces acting on the particles, leading to particle detachment and redistribution within the suspension. Under laminar flow conditions, this competition is commonly characterized by the Shields number, $\tau^* = \tau/(\Delta\rho g d)$, where $\tau$ is the applied shear stress, $\Delta\rho$ is the density difference between particles and fluid, $g$ is gravitational acceleration, and $d$ is the particle diameter[9,10]. Resuspension is typically observed when the applied stress becomes comparable to the gravitational stress scale, corresponding to $\tau^* = O(1)$[11–13].

Resuspension under shear flow has been extensively investigated through experimental and numerical studies aimed at understanding how sedimented particles become re-entrained into flowing suspensions[14–17]. Early work established that, for suspensions of spherical particles under steady shear, the equilibrium height of the resuspended layer results from a balance between gravitational settling and shear-induced particle migration, linking resuspension dynamics to the applied stress through the Shields number[7,8]. Subsequent studies showed that particle properties, including size, density, and polydispersity, influence migration rates and steady-state concentration profiles[18–20].

More recently, our prior work[21] demonstrated, through combined rheometry and local measurements, that resuspension dynamics are governed not only by the instantaneous shear rate but also by the accumulated deformation imposed on the suspension. In particular, we showed that a strain-based framework can unify resuspension behavior under both steady and oscillatory shear, revealing that particle mobilization is strongly controlled by flow history and collective particle rearrangements rather than stress magnitude alone[21,22]. These results further indicate that resuspension proceeds through a continuous transition rather than a sharp threshold, emphasizing the role of microstructural evolution in determining suspension states.

In parallel, extensive work has established that particle shape and anisotropy fundamentally influence the rheology of dense non-Brownian suspensions. Deviations from spherical geometry modify viscosity, shear thinning and thickening behavior, jamming thresholds, and stress transmission through changes in particle packing, excluded volume, and orientational dynamics[23–25]. Elongated and angular particles enhance frictional interactions, reduce maximum packing fractions, and broaden shear-jamming regimes, leading to stronger shear thickening and pronounced history-dependent hysteresis compared with spherical systems[26,27]. In addition to rheological effects, recent experimental and numerical studies indicate that particle anisotropy also alters sediment structure, migration behavior, and transport dynamics in sheared and sedimenting suspensions[28,29]. These findings highlight that particle morphology plays a central role in governing both microstructural evolution and macroscopic flow behavior, suggesting that its impact on resuspension dynamics remains an important open question.

Despite substantial advances in the rheology of anisotropic suspensions, the role of particle shape in viscous resuspension remains largely unexplored. Existing studies have predominantly focused on spherical particles, and it is unclear whether the recently established strain-based framework for dense suspensions extends to systems with particle anisotropy. Because elongated particles exhibit rotational dynamics, alignment, and tumbling in addition to translational migration, particle shape may fundamentally alter both the onset of resuspension and the subsequent pathway toward a fully suspended state.

Building on our recent work on spherical suspensions[21], which demonstrated that resuspension thresholds are governed by accumulated strain and can be organized within a unified framework, the present study investigates how particle anisotropy modifies resuspension dynamics under both steady and oscillatory shear. We perform experiments on dense suspensions composed of spherical and rod-shaped non-Brownian particles, systematically varying particle shape and volume fraction while combining rheological measurements with direct optical visualization. Our results show that the onset of resuspension occurs at a nearly identical critical strain for both particle shapes, indicating a morphology-independent criterion for particle detachment governed primarily by hydrodynamic forcing and gravitational confinement. In contrast, the strain required to reach a fully suspended state depends strongly on particle shape, with rod-shaped particles exhibiting delayed equilibration that is consistent with additional orientational, rotational, and steric constraints. These findings reveal that viscous resuspension proceeds through two distinct rheological transitions: an initial, shape-independent detachment process followed by a shape-dependent pathway toward full suspension strongly influenced by particle anisotropy and the additional structural rearrangements required for rod-shaped particles.

Overall, this work extends strain-based descriptions of viscous resuspension beyond spherical systems and establishes a direct link between particle anisotropy, rotational mobility, and suspension evolution under shear.

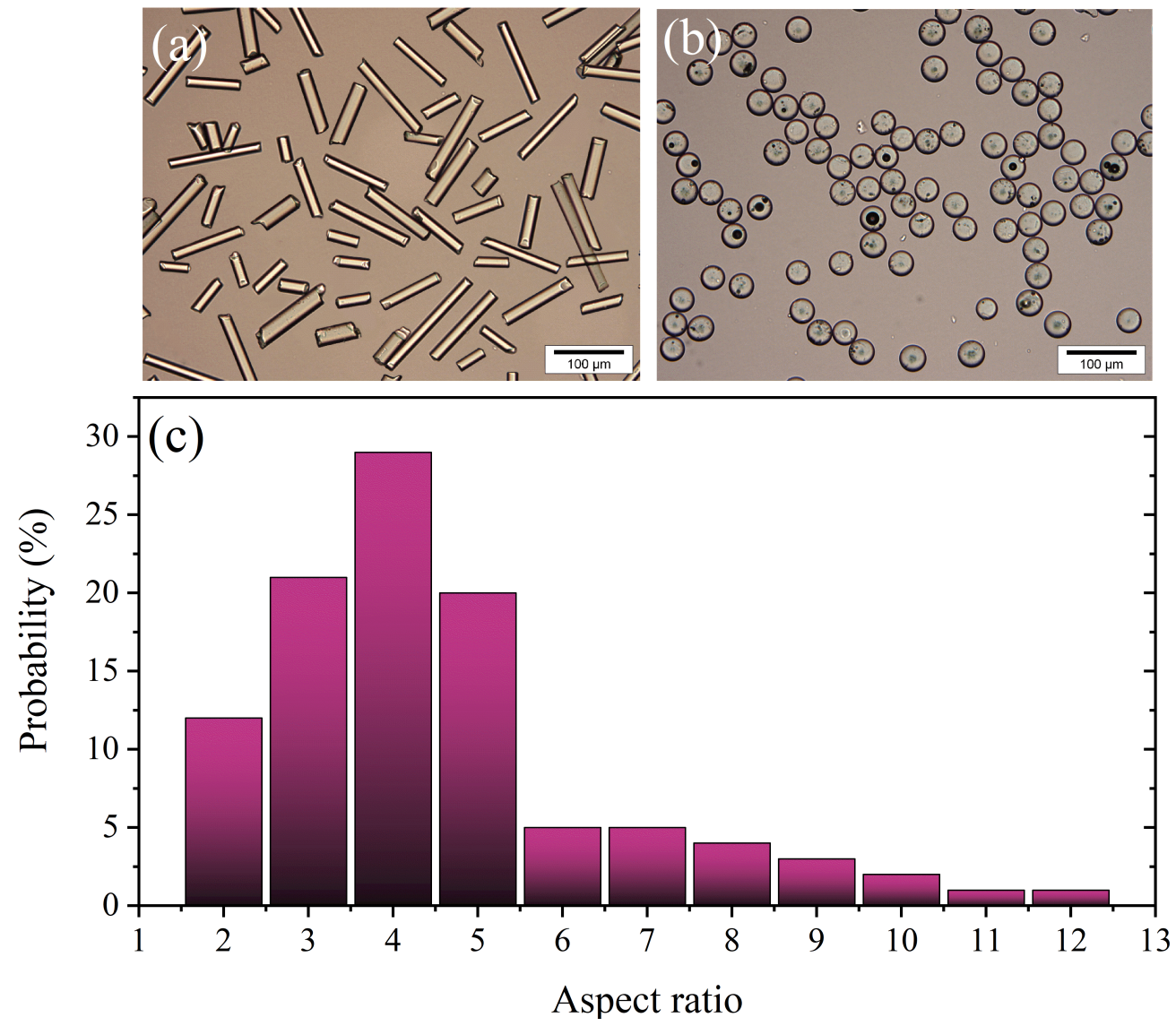

FIG. 1. Optical microscopy images of (a) glass fibers with an average length of $85 \pm 20$ $\mu$m and (b) glass spheres with an average diameter of $35 \pm 3$ $\mu$m. The scale bar corresponds to 100 $\mu$m. (c) Aspect ratio distribution of the glass fibers, with a mean aspect ratio of $\Gamma = 4.6$.

## II. MATERIALS AND METHODS

### A. Particles and suspension preparation

To isolate the effect of particle shape while minimizing variations due to material properties, two particle geometries—spherical particles and short rods—were selected from the same base material, glass, with a density of 2.5 $\mathrm{g\,cm^{-3}}$. The rod-shaped particles were produced from milled glass fibers (1/32 in., Fibreglast) and processed to obtain an average length of $85 \pm 20$ $\mu$m and a mean aspect ratio of 4.6 following a separation procedure based on sedimentation time. The spherical particles were obtained from Cospheric and had diameters in the range of 32–38 $\mu$m. This size range was selected such that the individual particle mass matched that of the glass rods, enabling a direct and controlled comparison of particle shape effects on flow behavior. Table I represents the geometric dimensions of each particle shape used in this study.

To assess whether elastic deformation of the fiber particles could influence the observed dynamics, the dimensionless bending stiffness was estimated following the formulation of Klingenberg *et al.*[30] as

$$\hat{K} = \frac{EI}{\eta_m \dot{\gamma} L^4}, \tag{1}$$

where $E$ is the Young's modulus of the glass fibers ($E \approx$ 70 GPa), $I = \pi d^4/64$ is the second moment of area for a cylindrical rod of diameter $d$, $\eta_m$ is the suspending fluid viscosity, and $L$ is the particle length. Using the maximum imposed shear rate ($\dot{\gamma} = 250$ s$^{-1}$), we obtain $\hat{K} \approx 1.4 \times 10^6$, with an estimated range of $3 \times 10^5$–$5 \times 10^6$ accounting for particle size variability. Because $\hat{K} \gg 1$, the rods behave as effectively rigid particles under the present conditions. Consequently, any observed differences relative to spheres can therefore be attributed to geometry and orientation dynamics rather than elastic deformation.

Figure 1(a,b) present optical microscopy images of the glass rods and spheres, respectively, while Figure 1(c) shows the aspect-ratio distribution of the rod-shaped particles (mean $\Gamma = 4.6$). The rod dimensions were quantified through image analysis of more than 500 individual particles using ImageJ software. The particles were suspended in AR20 silicone oil

TABLE I. Particle shapes and geometric dimensions used in the experiments.

| Shape | Length ($\mu$m) | Diameter ($\mu$m) | Aspect Ratio |
|---|---|---|---|
| Rod | $85 \pm 20$ | $18 \pm 2$ | 4.6 |
| Sphere | - | $35 \pm 3$ | 1 |

(density $\rho_m = 1.00\,\mathrm{g\,cm^{-3}}$, Sigma-Aldrich), which behaves as a Newtonian fluid with a dynamic viscosity of $\eta_m = 0.02$ Pa s. Owing to the density mismatch between the particles and the suspending medium, gravitational forces act on the particles, resulting in sedimentation within the suspension.

Suspensions were prepared by dispersing the particles in the silicone oil using vortex mixing for 60 s. A range of particle volume fractions was investigated for both spherical and rod-shaped particles (see Table II for details). The particle volume fraction, $\phi$, was calculated from the particle mass fraction via[31]

$$\phi = \frac{\rho_m x}{\rho_m x + \rho_d (1-x)}, \tag{2}$$

where $\rho_m$, $\rho_d$, and $x$ denote the liquid density, particle density, and particle mass fraction, respectively.

### B. Rheological measurements

Rheological characterization was performed using a DHR-2 rotational rheometer (TA Instruments) equipped with a parallel-plate geometry of 50 mm diameter. The rheometer gap was fixed at 1.2 mm, corresponding to approximately 14 times the average rod length, which is sufficient to suppress finite-size and confinement effects. All measurements were carried out at a controlled temperature of 20 °C, at which the silicone oil exhibits a viscosity of $\eta_m = 0.02$ Pa s.

To mitigate wall slip, sandpaper was affixed to both plates; comparative tests with smooth plates showed no measurable differences in the rheological response.

Following the preshear and sedimentation protocol established for spherical suspensions[21], each sample was presheared at a constant shear stress of $\tau = 15$ Pa for 120 s to erase the initial particle structures. Samples were then allowed to rest under zero applied stress for controlled durations (0–300 s) to promote sedimentation. Oscillatory shear experiments were conducted with zero conditioning time and three oscillation cycles per data point; the negligible influence of cycle number and frequency on resuspension is documented in our recent work[21].

### C. Optical imaging and microstructure visualization

To visualize the flow field and the evolution of particle microstructure during resuspension, high-speed imaging was performed within the rheometer gap. A Phantom MIRO LAB 320 camera, equipped with a 2× F-mount adapter and a Navitar 12× adjustable zoom lens, was used for image acquisition. Images were recorded at a resolution of 1920 × 1200 pixels. Illumination was provided by a 150 W standalone LED light source (MultiLed QT, GSVITEC), ensuring uniform lighting during high-speed capture.

Particle geometry and dimensions were characterized using optical microscopy, followed by ImageJ-based postprocessing. Representative images are shown in Figure 1.

## III. RESULTS

### A. Maximum packing fraction of particles

Particle resuspension between parallel plates is strongly influenced by the height of the sediment bed. To accurately determine this height for a given particle volume fraction, it is essential to quantify the maximum packing fraction, $\phi_j$, for each particle shape. Previous studies have shown that deviations from spherical geometry alter $\phi_j$, which typically decreases with increasing particle aspect ratio[23–26]. Building on these findings, and our recent work on spherical suspensions[21], we first characterize $\phi_j$ for both spherical and rod-like particles to establish a consistent basis for comparing resuspension behavior across particle shapes.

To this end, we measure the relative viscosity, $\eta_r$, of the suspensions over a range of particle volume fractions spanning moderate to highly concentrated regimes. The viscosity measurements are performed under a constant shear stress of 15 Pa for 300 s, conditions chosen to ensure that the suspensions reach a steady, fully dispersed state. This approach allows us to evaluate $\eta_r$ under fully suspended regime conditions and reliably determine $\phi_j$ for each particle system.

Figure 2 shows the relative viscosity as a function of particle volume fraction for rod-shaped (squares) and spherical (circles) suspensions. The maximum packing fraction, $\phi_j$, is estimated by fitting the data to the Krieger–Dougherty model[32], which describes the divergence of viscosity as $\phi \to \phi_j$. The fitted model is used to extrapolate the divergence point, which is taken as the maximum packing fraction for each particle shape. In this framework, the relative viscosity is expressed as

$$\eta_r = k\left(1 - \frac{\phi}{\phi_j}\right)^{-\beta}, \tag{3}$$

where $\eta_r$ denotes the suspension viscosity normalized by the viscosity of the suspending fluid, while $\phi_j$ represents the jamming, or maximum packing, fraction. The parameter $\beta$ is a fitting constant and is commonly taken to be approximately equal to 2 for non-Brownian suspensions[25,33–35]. The solid curves in Figure 2 show the fits of the Krieger–Dougherty model to the experimental data.

The divergence of the relative viscosity (dashed lines in Figure 2) is used to identify the maximum packing fraction, yielding values of $\phi_j = 0.475$ for rod suspensions and $\phi_j =$

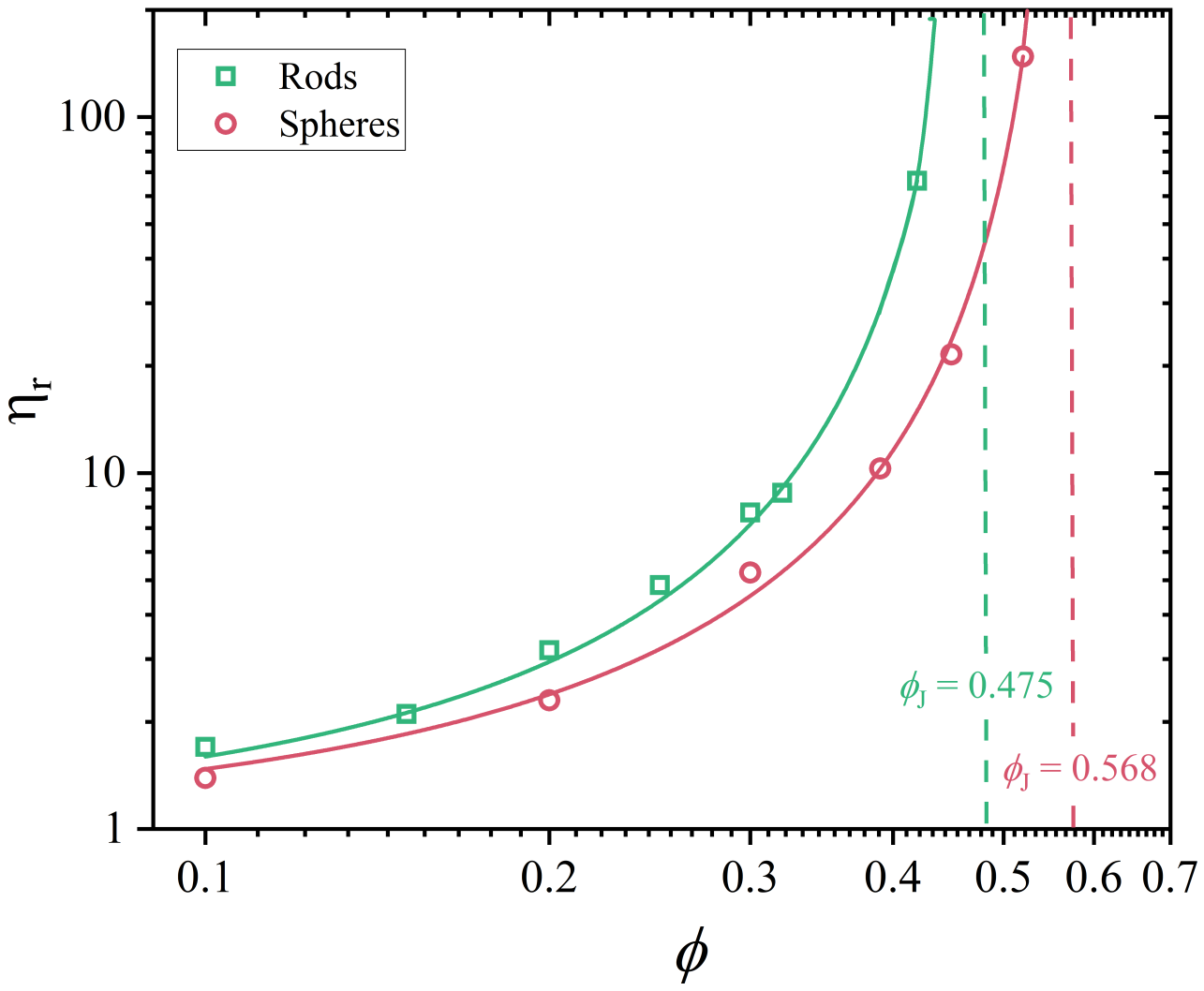

FIG. 2. Relative viscosity as a function of particle volume fraction for rod-shaped particle suspensions (squares) and spherical particle suspensions (circles). Solid lines represent fits to the Krieger–Dougherty model (Eq.8), showing extrapolated jamming volume fractions of $\phi_j = 0.475$ for rods and $\phi_j = 0.568$ for spheres, respectively.

TABLE II. Suspension concentrations and corresponding normalized volume fractions used in rheological measurements.

| $\phi_{Sphere}$ | $\phi_{Rod}$ | $\bar{\phi} = \phi/\phi_j$ |
|---|---|---|
| 0.32 | 0.27 | 0.56 |
| 0.39 | 0.32 | 0.68 |
| 0.45 | 0.38 | 0.79 |
| 0.53 | 0.42 | 0.90 |

0.568 for sphere suspensions. The lower packing fraction observed for the rod-based suspensions is consistent with previous studies, which have demonstrated that increasing particle aspect ratio and angularity reduces packing efficiency in non-Brownian suspensions[23–26]. This implies that, at the same particle volume fraction, rod-shaped particles collectively occupy a larger effective volume than spherical particles. Consequently, in the sedimented state, rod suspensions form a taller sediment bed compared to their spherical counterparts. Figure 3 schematically illustrates the difference in maximum packing volume and resulting sediment height for suspensions of spheres and rods at identical solid volume fractions. To ensure a consistent basis for comparing resuspension behavior across different particle shapes, all suspension systems were therefore characterized using an equal normalized volume fraction $\bar{\phi} = \phi/\phi_j$, where $\phi$ is the particle volume fraction and $\phi_j$ is the corresponding jamming packing fraction. The suspension compositions used in the resuspension rheological experiments are summarized in Table II, where four distinct normalized volume fractions are defined for both particle shapes.

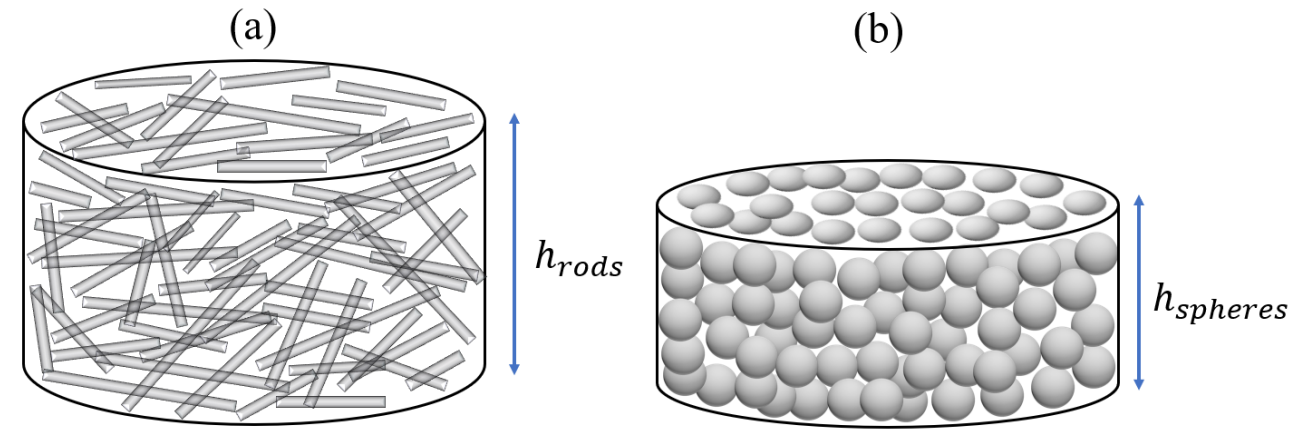

FIG. 3. Schematic comparing particle packing volume and resulting sediment height at the same solid volume fraction for (a) rod-shaped particles and (b) spherical particles, highlighting the effect of particle shape on sediment structure.

## B. Resuspension Dynamics Under Steady Shear

To investigate the influence of particle shape on transient resuspension dynamics, steady-shear experiments were performed on suspensions of spherical and rod-shaped particles at different normalized volume fractions. Following the protocol established in our previous study[21], each experiment consisted of a 120 s preshear to remove pre-existing microstructure, followed by 300 s of quiescent sedimentation under zero applied stress. This sedimentation period exceeds the characteristic settling time of the suspension ($\sim 100$ s) and ensures the formation of a reproducible sediment bed before shearing. Resuspension was then initiated by abruptly imposing a constant shear rate of $\dot{\gamma} = 50\ \mathrm{s}^{-1}$. To confirm that the experiments were conducted in the viscously dominated regime, we estimate the Bagnold number for the maximum shear rate,

$$Ba = \frac{\rho_p d^2 \dot{\gamma}}{\eta_m}, \tag{4}$$

where $\rho_p$ is the particle density, $d$ is the characteristic particle size, and $\eta_m$ is the suspending-fluid viscosity. For the representative experimental conditions considered here, the estimated Bagnold number is $Ba \approx 7 \times 10^{-3}$. This value is well below the transition to grain-inertia-dominated transport, indicating that particle motion is governed primarily by viscous hydrodynamic interactions and particle contacts rather than collisional inertia. Consequently, the resuspension dynamics are expected to be controlled by microstructural rearrangements and accumulated deformation.

Figure 4 presents the transient evolution of the shear stress during resuspension for suspensions with normalized volume fractions of $\bar{\phi} = 0.56$, 0.68, and 0.90. For both particle shapes, the stress evolution reveals three distinct regimes corresponding to different suspension microstructures.

In *Regime I (bed-load transport)*, deformation is accommodated primarily within the clear fluid layer above the sediment bed, while most particles remain confined within a mechanically stable contact network. As a result, only weak stress growth is observed[11,13,36]. In *Regime II (resuspension)*, progressive particle entrainment activates hydrodynamic interactions and force transmission throughout the bed, producing a rapid increase in stress as an increasing fraction of particles becomes mobile. Finally, in *Regime III (fully suspended*

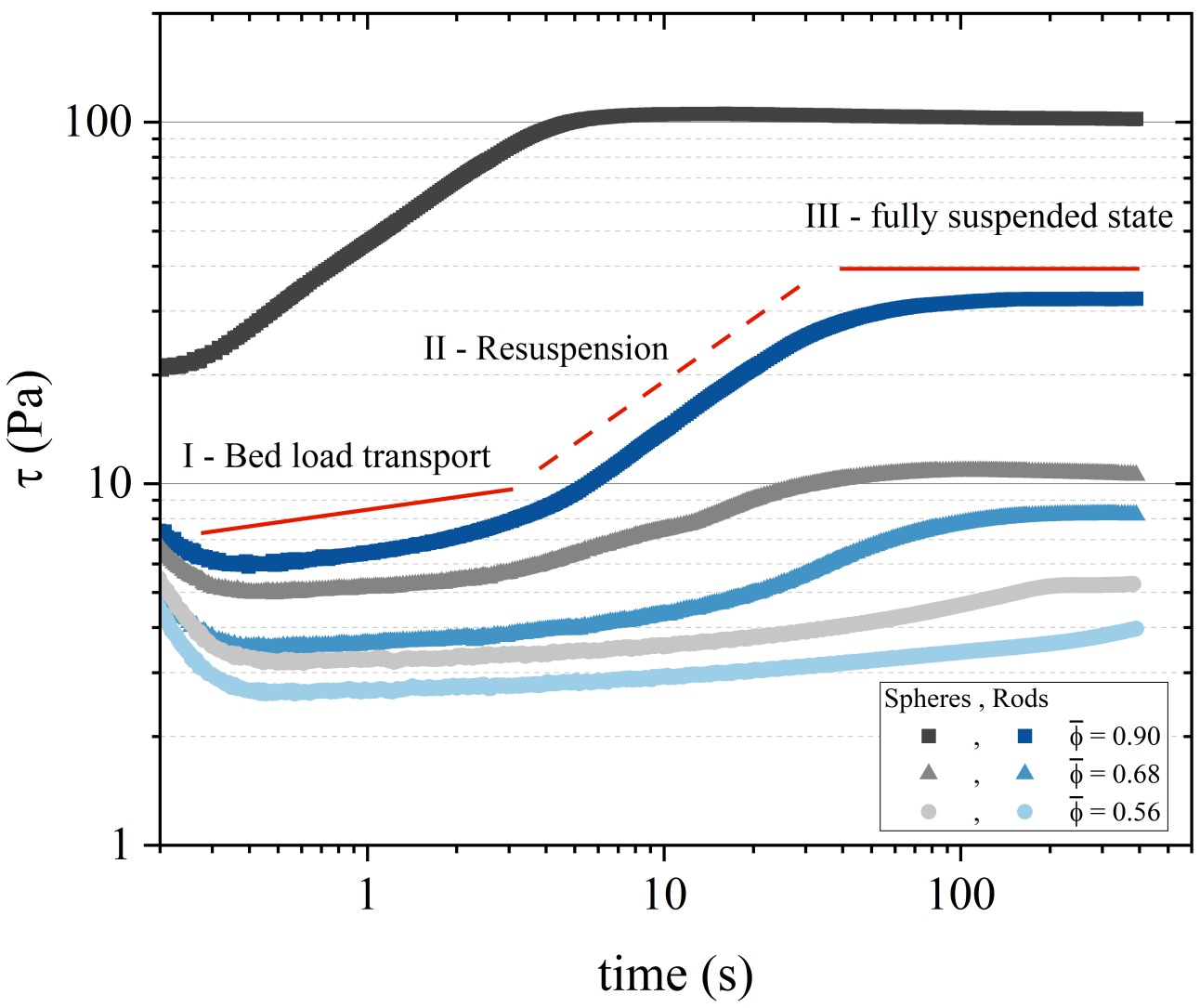

FIG. 4. Transient evolution of shear stress under steady shear at $\dot{\gamma} = 50\ s^{-1}$. Shear stress is plotted as a function of strain for suspensions with normalized volume fractions $\bar{\phi} = 0.56$, 0.68, and 0.90, for both spherical particles (shades of gray) and rod-shaped particles (shades of blue). Red lines indicate the three resuspension regimes: (I) bed-load transport, (II) resuspension, and (III) fully suspended state.

*state)*, particle entrainment and settling reach a statistical balance, resulting in a dynamically equilibrated suspension microstructure and a steady rheological response.

A comparison between spherical and rod-shaped suspensions reveals that the transition to the fully suspended state occurs systematically later for rods at all normalized volume fractions. This observation demonstrates that particle shape significantly influences the transient pathway toward fully suspended regime. Similar trends have been reported in anisotropic particle and fiber suspensions, where rotational constraints and particle alignment modify transient rheological behavior and delay structural equilibration[23,24,27,37].

The observations suggest that rotational reorganization contributes to the delayed transition. Unlike spherical particles, rods must undergo substantial rotational reorganization and alignment before becoming fully entrained within the flow. Consequently, resuspension involves not only particle detachment and migration but also progressive rotational relaxation, introducing an additional stage of microstructural evolution that is absent in spherical systems. The prolonged transient response may therefore be interpreted as a slower structural reformation process, in which translational migration and rotational rearrangement evolve simultaneously before a fully suspended state can be established.

Figure 5 illustrates the corresponding evolution of the suspension microstructure during resuspension. As shear penetrates into the sediment bed, particles become progressively mobilized, resulting in bed expansion, particle migration, and eventual fully suspended regime of the suspension. For rod-shaped particles, this process is accompanied by alignment and rotational reorganization, which contribute to the slower evolution toward the fully suspended state.

To establish a unified framework for comparing resuspension behavior across particle shapes and concentrations, strain-controlled flow-ramp experiments were subsequently performed using the same preshear and sedimentation protocol. Figures 6(a) and 6(b) show the evolution of the apparent viscosity as a function of accumulated strain for rod-shaped and spherical suspensions, respectively.

At small strains, the viscosity remains approximately constant, reflecting a sediment-dominated microstructure in which deformation is accommodated primarily within the interstitial fluid layer above the bed. As strain accumulates, a pronounced increase in viscosity is observed for all suspensions. This increase does not simply reflect a greater number of particles participating in the flow; rather, it signifies the progressive development of hydrodynamic interactions, particle contacts, and stress-transmission pathways throughout the suspension. The resulting viscosity growth therefore provides a rheological measure of the evolving suspension microstructure during resuspension. Importantly, the transition from sedimented to fully suspended states occurs over a finite strain interval rather than at a single critical strain. This behavior indicates that resuspension is governed by cumulative microstructural rearrangements and progressive structural evolution rather than by an instantaneous force balance alone. Similar strain-controlled transitions have been reported in dense non-Brownian suspensions, where accumulated deformation governs particle rearrangement and migration processes more strongly than the instantaneous shear rate[38–40].

A systematic dependence on particle concentration is observed, with the viscosity transition shifting to larger strains as the normalized volume fraction increases. This trend indicates that denser suspensions require more extensive microstructural reorganization before reaching a fully suspended state. The influence of particle shape is particularly evident during the later stages of the transition, where the system reaches a fully dispersed stage. While the onset of viscosity growth occurs at comparable strain levels for rods and spheres, the subsequent approach to the fully suspended state is substantially delayed for rod suspensions. This extended transition reflects the additional orientational and steric constraints associated with anisotropic particles. Thus, the steady-shear results demonstrate that viscous resuspension is fundamentally a strain-controlled process governed by microstructural evolution. Particle shape has only a modest influence on the initial stages of mobilization but strongly affects the subsequent structural evolution required to achieve a fully suspended regime. Rod-shaped particles, therefore, require larger accumulated deformation to reach a fully suspended state, likely because translational migration is accompanied by rotational reorganization and alignment.

### C. Resuspension Dynamics Under Oscillatory Shear

To investigate the influence of oscillatory deformation on viscous resuspension, a series of oscillatory shear experiments was performed by systematically varying the strain amplitude, particle concentration, and sedimentation time. Build-ing upon the strain-controlled framework recently established for spherical suspensions[21], the objective here is to determine whether the rheological transitions associated with resuspension remain governed by accumulated deformation in the presence of particle anisotropy.

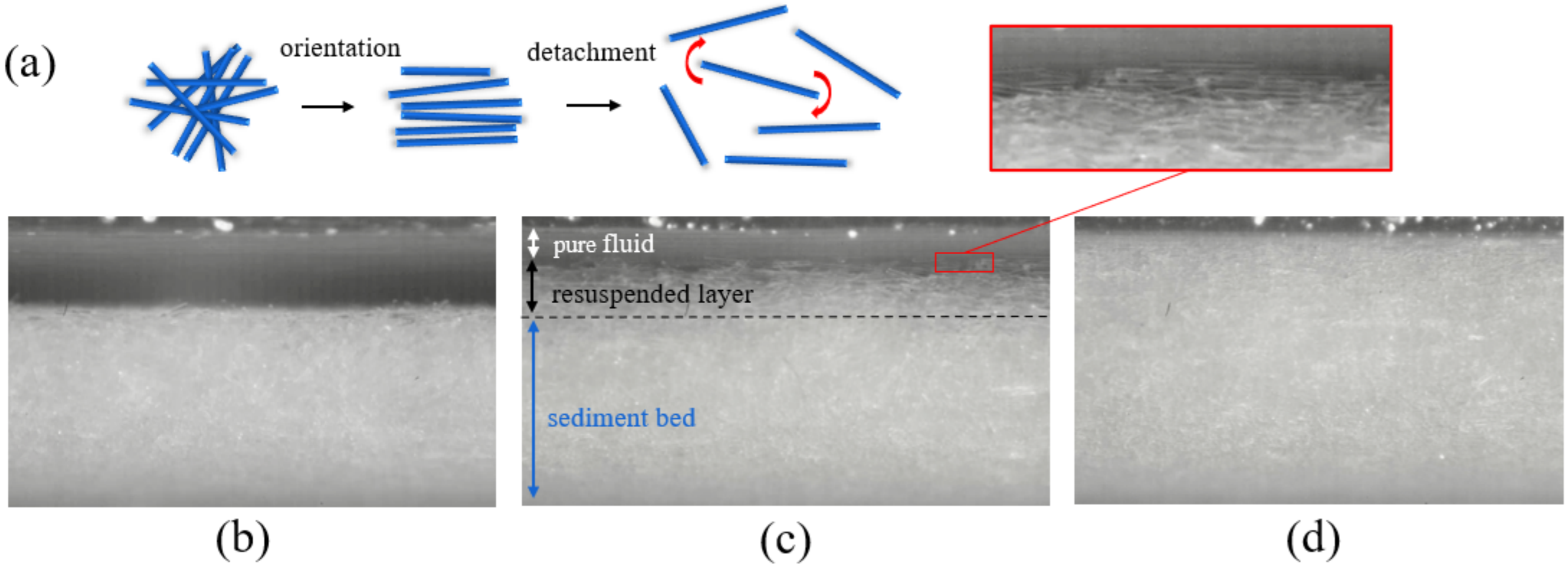

FIG. 5. Evolution of particle microstructure during shear-induced resuspension. (a) Schematic depicting the alignment, rotation, and detachment of rod-shaped particles in the resuspended layer. (b–d) Side-view images of the rheometer gap illustrating particle distributions in the (b) sedimented state, (c) resuspension regime, and (d) fully dispersed suspension.

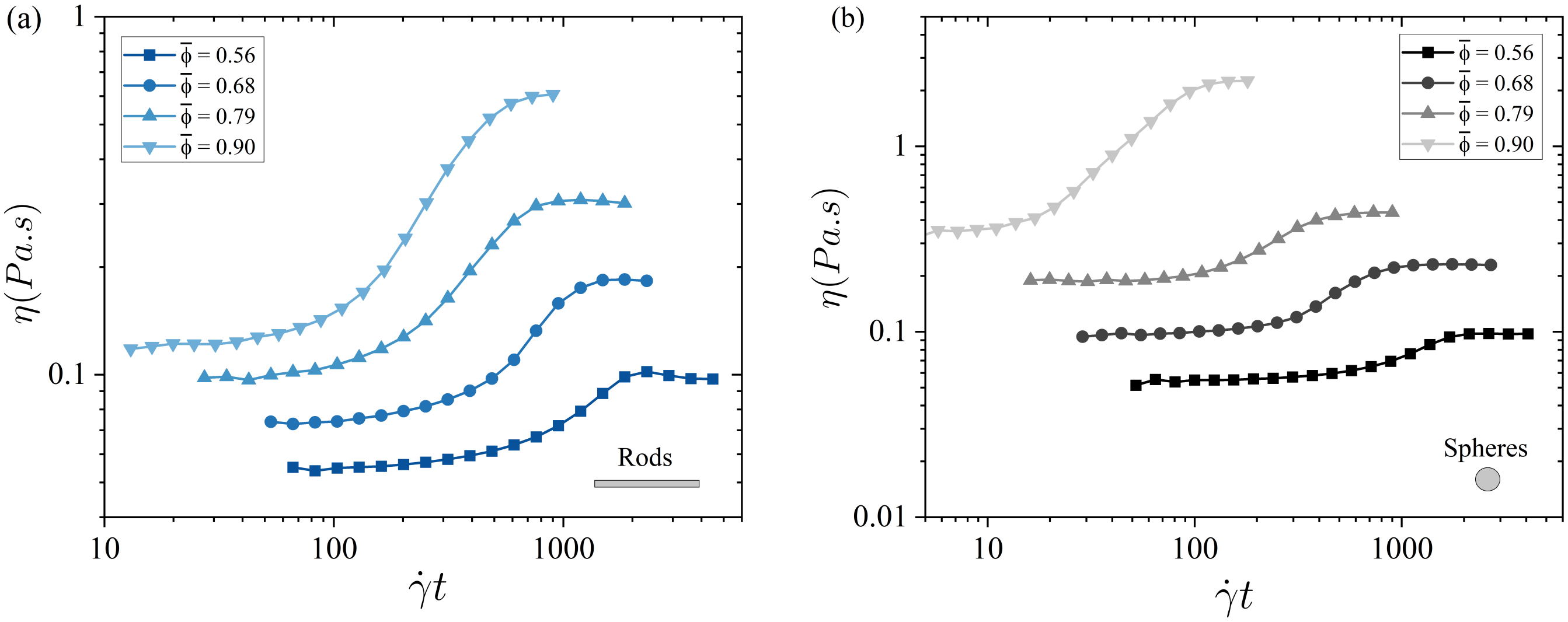

FIG. 6. Comparison of shear-induced resuspension under steady shear for normalized particle volume fractions ranging from $\bar{\phi} = 0.56$ to 0.90, obtained from flow ramp experiments. Viscosity $\eta$ is plotted as a function of cumulative strain for suspensions of (a) rod-shaped particles and (b) spherical particles.

Suspensions of glass spheres and glass rods dispersed in a Newtonian silicone oil were first presheared at 15 Pa for 120 s to eliminate flow-history effects and establish a reproducible initial microstructure. Following preshear, the suspensions were allowed to rest for 0–300 s to achieve different degrees of sedimentation before oscillatory forcing was applied. A sinusoidal shear strain, $\gamma(t) = \gamma_g \sin(\omega t)$ was then applied using a parallel-plate geometry. Figure 7 presents the magnitude of the complex viscosity, $|\eta^*|$, as a function of the global strain amplitude for spherical and rod suspensions at $\bar{\phi} = 0.68$. Similar to the transient behavior observed under steady shear, the oscillatory response exhibits three rheological states corresponding to sedimented, partially suspended, and fully suspended microstructures. Two critical strain amplitudes can be identified. The first, $\gamma_{g,c1}$, marks the onset of particle mobilization and the breakdown of the sediment structure. The second, $\gamma_{g,c2}$, corresponds to the completion of resuspension and the establishment of a dynamically equilibrated suspension state. Beyond this threshold, the measured viscosity becomes independent of the initial sediment configuration, indicating that particle redistribution and sedimenta-tion have reached a dynamic balance over each oscillatory cycle.

The existence of two distinct critical strain amplitudes demonstrates that viscous resuspension is not an instantaneous event but rather a finite strain-controlled transition extending from initial particle detachment to a fully suspended regime. Importantly, both transition thresholds remain nearly invariant with respect to sedimentation time. Despite substantial differences in the degree of settling and sediment structure, the same critical strain amplitudes are recovered over the entire range of rest times investigated. The observed variation remains within experimental uncertainty, indicating that the transition thresholds are robust and reproducible.

This collapse demonstrates that accumulated deformation, rather than the details of the initial sediment structure, serves as the primary control parameter governing the rheological transition. Similar strain-controlled behavior has been reported in periodically driven suspensions exhibiting reversible-to-irreversible transitions, where cumulative deformation governs particle rearrangements more strongly than the instantaneous forcing conditions[22,38,39]. The invariance of the critical strain despite substantial differences in sediment structure further suggests that particle mobilization is governed by a local deformation criterion rather than the global packing configuration. Although sedimentation modifies the thickness and concentration profile of the packed bed, the onset of resuspension occurs when particles near the sediment–fluid interface experience a sufficiently large local fluid strain to initiate irreversible rearrangements. Consequently, the critical strain represents a robust local rheological condition that is largely independent of the initial sediment structure.

While both particle shapes exhibit rest-time-independent transition thresholds, the strain interval separating the onset and completion of resuspension is consistently larger for rod suspensions than for spherical suspensions. This observation indicates that particle anisotropy has only a limited influence on the initiation of particle mobilization but strongly affects the subsequent evolution toward a fully suspended regime. The larger strain interval required for rods is consistent with the orientational dynamics discussed in steady shear results (Section III.B), suggesting that rotational reorganization contributes primarily to the later stages of full suspension rather than the onset of particle detachment.

To further investigate the temporal evolution of these rheological states, time-sweep experiments were performed at fixed strain amplitudes ranging from 10% to 5000% following preshear and a 300 s sedimentation period. Figure 8 shows the corresponding evolution of the complex viscosity for a representative rod suspension.

Two qualitatively distinct responses are observed. At strain amplitudes below the first critical threshold, the viscosity remains nearly constant throughout the experiment, indicating that the sedimented microstructure remains largely intact and only limited particle rearrangement occurs. In contrast, strain amplitudes exceeding $\gamma_{g,c1}$ produce a pronounced increase in viscosity followed by a steady-state plateau. The transient growth period reflects progressive particle entrainment, increasing hydrodynamic coupling, and the development of suspension-scale stress-transmission pathways. The eventual plateau indicates that particle redistribution and sedimentation have reached a dynamic balance, resulting in a statistically stationary suspension microstructure.

The results demonstrate that the characteristic timescale governing oscillatory resuspension is more naturally expressed in terms of accumulated strain than elapsed time.

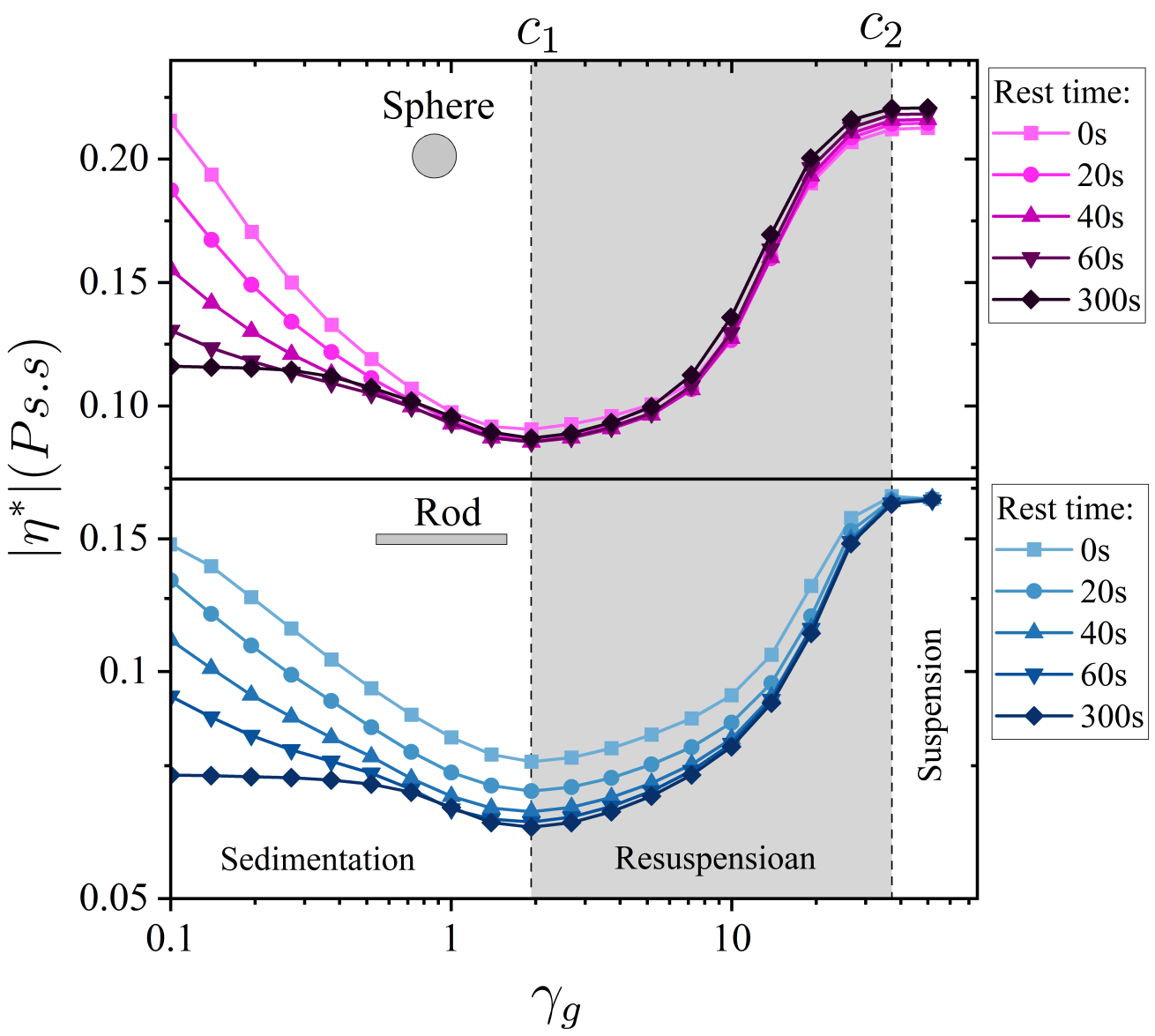

FIG. 7. Magnitude of the complex viscosity, $|\eta^*|$, as a function of global strain amplitude for different rest times (0–300 s) at $\bar{\phi} = 0.57$. The upper and lower panels correspond to suspensions of spherical and rod-shaped particles, respectively. The first and second global critical strains, $\gamma_{g,c1}$ and $\gamma_{g,c2}$, denote the onset and completion of the resuspension process.

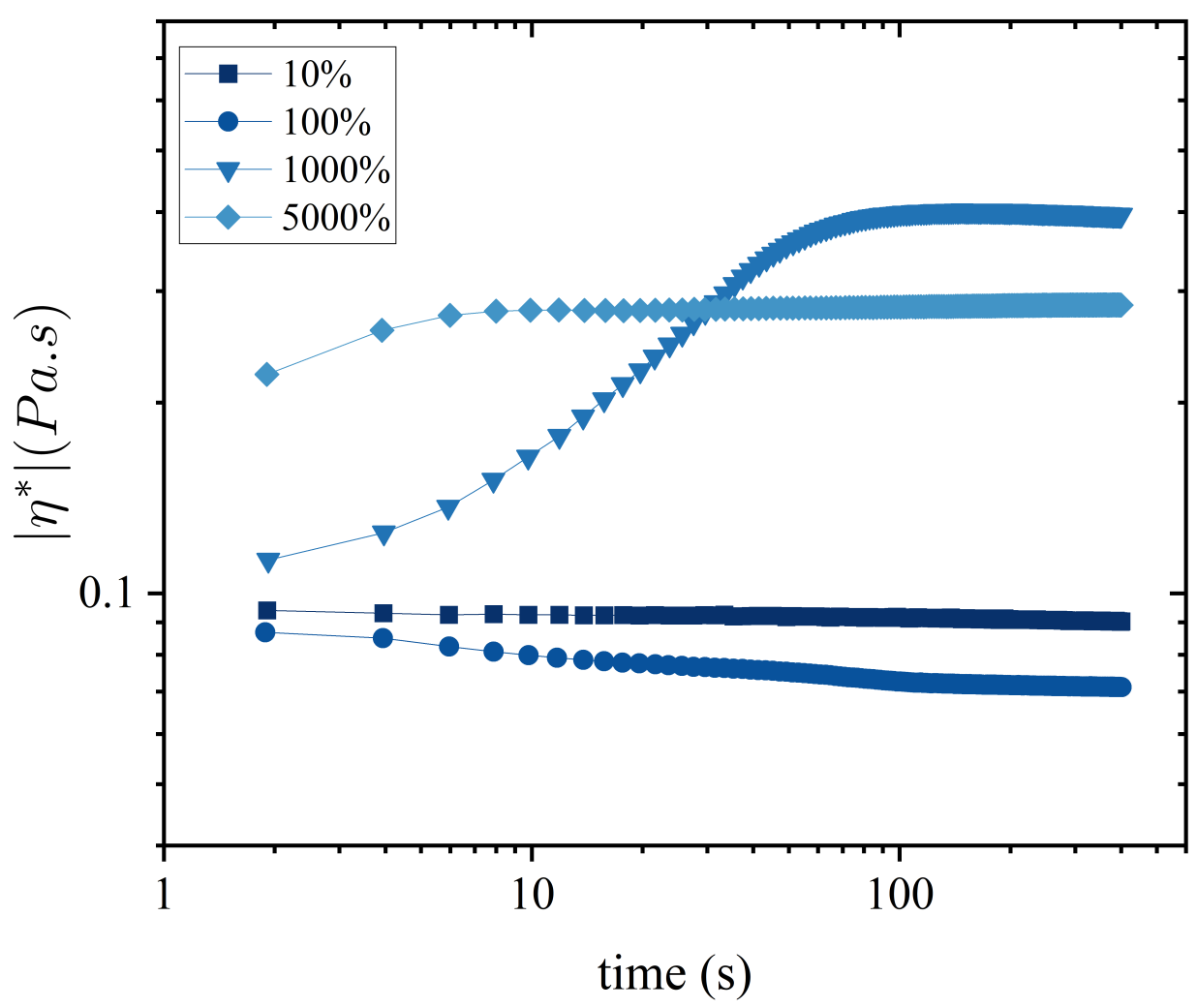

FIG. 8. Temporal evolution of the magnitude of the complex viscosity, $|\eta^*|$, during continuous oscillatory shear for a rod-shaped particle suspension at $\bar{\phi} = 0.58$ and $\omega = 10$ rad/s. Data are shown for strain amplitudes of 10%, 100%, 1000%, and 5000%, highlighting the strain-dependent response of the suspension.

When viewed within this strain-based framework, the rheological evolution collapses onto two well-defined transitions corresponding to particle mobilization and fully suspended regime. Figures 7 and 8 therefore demonstrate that oscillatory resuspension is governed by two distinct rheological transitions: the onset of particle detachment from the sediment bed and the subsequent establishment of a dynamically equilibrated fully suspended state. These observations provide the foundation for the unified strain-scaling analysis presented in the following section.

## IV. DISCUSSION

To systematically examine the influence of particle concentration and particle shape on resuspension dynamics, oscillatory shear experiments were performed over a range of normalized volume fractions, $\bar{\phi} = 0.56$–$0.90$. The magnitude of the complex viscosity, $|\eta^*|$, was measured as a function of increasing strain amplitude following complete sedimentation of the suspension. Because higher particle loadings produce larger sediment-bed heights, the imposed global strain does not directly represent the deformation experienced by the particle packing. A central objective of this study is therefore to determine whether a unified strain-based description of resuspension exists across different particle concentrations and morphologies.

Figure 9(a) presents the evolution of the complex viscosity as a function of global strain amplitude for spherical and rod-shaped suspensions. At low strain amplitudes, the imposed deformation is insufficient to overcome gravitational confinement and particle-contact forces, and the suspension remains sedimented, resulting in a plateau in $|\eta^*|$. As the strain amplitude increases, the viscosity rises sharply, indicating the onset of particle mobilization and resuspension. The corresponding transition is identified by the first critical global strain, $\gamma_{g,c1}$. Notably, $\gamma_{g,c1}$ decreases systematically with increasing particle concentration. This trend reflects the fact that denser suspensions possess thinner fluid layers above the sediment bed, causing a larger fraction of the imposed deformation to be transmitted directly to the particles.

To account for this geometric effect, we introduce the fluid strain, $\gamma_f$, which represents the local strain transmitted through the clear fluid layer to the sediment bed. For a parallel-plate geometry,

$$\gamma_f = \gamma_g \frac{h_{\text{gap}}}{h_f}, \tag{5}$$

where $h_{\text{gap}}$ is the total gap between the plates and $h_f$ is the thickness of the clear fluid layer above the sediment bed. The fluid layer thickness can be written as

$$h_f = h_{\text{gap}} - h_0 = h_{\text{gap}}(1 - \bar{\phi}), \tag{6}$$

where $h_0$ is the height of the sediment bed and $\bar{\phi} = h_0/h_{\text{gap}}$ is the normalized volume fraction. Substituting this expression into the strain relation yields the net strain applied to the particle packing,

$$\gamma_f = \frac{\gamma_g}{1 - \bar{\phi}}. \tag{7}$$

This scaling demonstrates that as the sediment bed occupies a larger fraction of the gap, the fluid layer becomes thinner and the local strain transmitted to the particles is amplified relative to the imposed global strain.

The fluid-strain framework reveals a remarkable collapse of the resuspension thresholds across particle concentrations and particle shapes. Figure 9(b) shows that the onset of resuspension occurs at an approximately constant critical fluid strain for both spherical and rod-shaped suspensions. The corresponding critical value, $\gamma_{f,c1} \approx 6$, is nearly independent of particle morphology, indicating that particle detachment is governed primarily by a local hydrodynamic lift-off criterion rather than by the detailed structure of the sediment bed. This result suggests that the initiation of resuspension is controlled by the local deformation experienced by particles at the sediment–fluid interface and is largely insensitive to particle shape.

A different behavior emerges as the system approaches the maximum packing condition, $\bar{\phi} = 0.90$. Under these highly confined conditions, the previously observed consistent collapse begins to break down. In particular, the onset strain for rod suspensions shifts to significantly larger values, indicating that greater accumulated deformation is required to initiate particle detachment. This increase in the resuspension threshold is consistent with the severe reduction in available free volume near the interface, which restricts particle reorientation and tumbling. Unlike more dilute conditions, where rods can rotate relatively freely, strong confinement imposes geometric constraints that hinder rotational motion. As a result, a particle must displace its neighbors to create sufficient space for tumbling, a process that requires additional strain to overcome local crowding and mechanical resistance[27,41].

The second resuspension threshold exhibits an even stronger dependence on particle morphology. For spherical suspensions, the completion of resuspension occurs at an approximately constant strain, $\gamma_{f,c2} \approx 120$, across all concentrations. In contrast, rod suspensions require substantially larger deformation, $\gamma_{f,c2} \approx 180$, to reach a fully suspended regime. This approximately 50% increase demonstrates that particle shape has a relatively minor influence on the initiation of particle detachment but strongly affects the subsequent evolution toward a fully suspended state.

For quantitative context, the rotational dynamics of an isolated ellipsoidal particle in unbounded simple shear can be described by Jeffery orbit theory, which relates the accumulated strain required for a full end-to-end tumbling event to the particle aspect ratio through a closed-form solution for periodic rotation[42–45]. The corresponding strain scale for one complete tumbling cycle ($2\pi$) is approximately five times larger than the first resuspension threshold measured in the present study. This comparison indicates that complete end-to-end tumbling is not required for the onset of particle mobilization. However, the dense and confined suspension considered here differs significantly from the dilute, unbounded conditions underlying

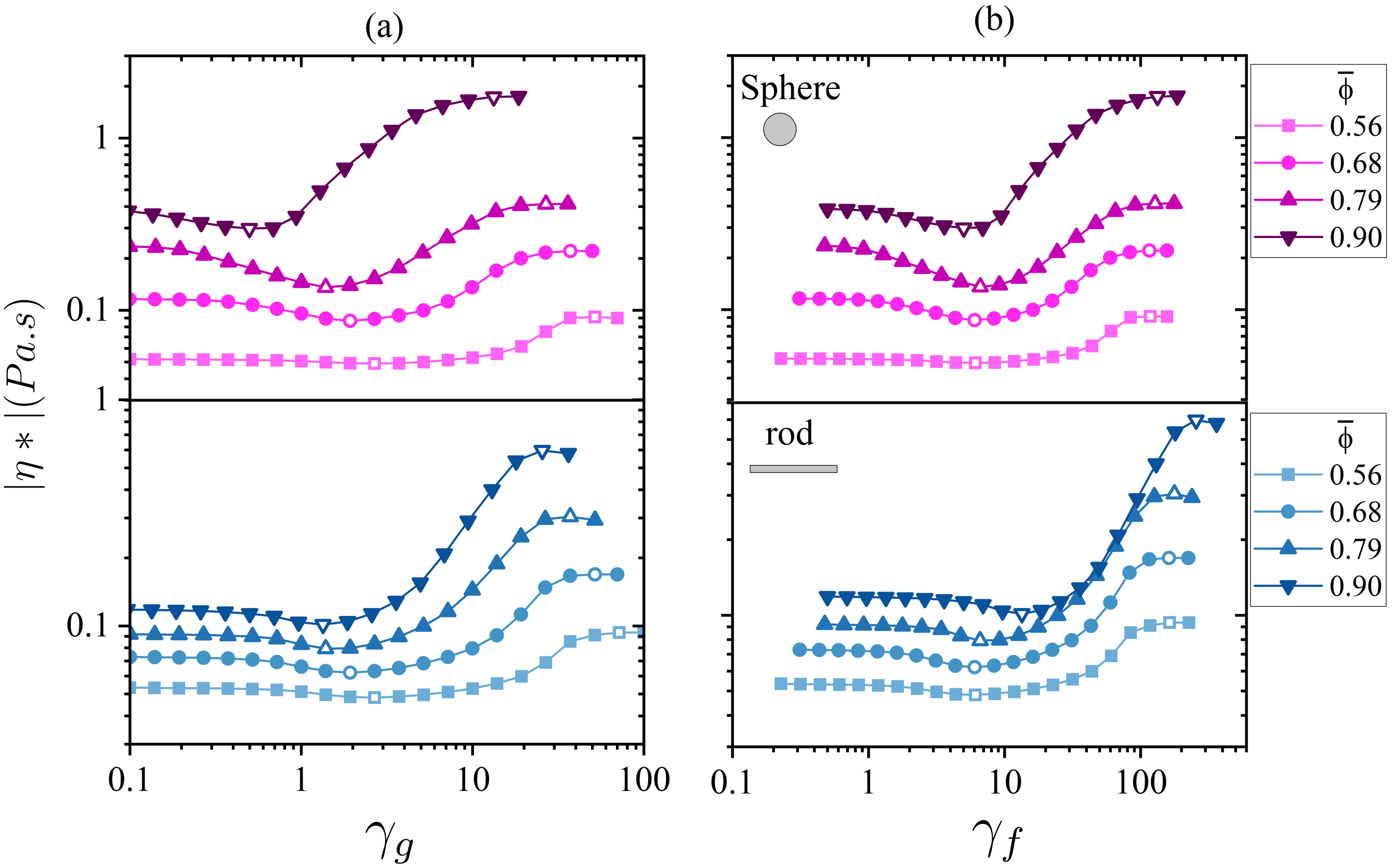

FIG. 9. Magnitude of the complex viscosity, $|\eta^*|$, as a function of (a) the global strain amplitude, $\gamma_g$, and (b) the local strain amplitude exerted by the fluid, $\gamma_f$, for normalized particle volume fractions ranging from $\bar{\phi} = 0.56$ to 0.90. The upper and lower panels correspond to suspensions of spherical and rod-shaped particles, respectively. Open symbols indicate the critical strain associated with the resuspension thresholds.

Jeffery's theory. Particle contacts, many-body hydrodynamic interactions, confinement, and local concentration gradients all modify the rotational dynamics of the rods. Therefore, Jeffery orbit calculations are used only to provide a qualitative rotational strain scale, rather than a quantitative prediction of the resuspension threshold. Additional details on Jeffery orbit kinematics and orientation evolution are provided in the Supplementary Material. It should be noted that although orientation was not quantified continuously throughout the resuspension process, the consistent increase in the strain required to reach the fully suspended state, together with established rotational dynamics of anisotropic particles under shear, supports the interpretation that orientational reorganization contributes primarily to the second resuspension transition rather than the onset of particle mobilization.

To summarize the resuspension behavior across a wide range of particle concentrations, Figure 10 presents state diagrams for both spherical and rod-shaped suspensions. The lower boundary denotes the onset of particle detachment from the sediment bed, while the upper boundary marks the transition to a fully suspended state. The region between these limits therefore defines a strain window over which partial resuspension occurs for each particle morphology.

These state diagrams reveal that viscous resuspension proceeds through two distinct rheological transitions. The first is a detachment transition, characterized by a nearly constant critical strain, $\gamma_{f,c1} \approx 6$, that is remarkably insensitive to particle shape over a broad range of volume fractions. This collapse suggests that the onset of resuspension is governed primarily by local hydrodynamic forcing at the sediment–fluid interface, rather than by particle-scale geometry. In contrast, the second transition—corresponding to the attainment of a fully suspended state—is controlled by the evolution and stability of the suspension microstructure and exhibits a strong dependence on particle anisotropy. Achieving this regime requires extensive particle rearrangement and sustained mobility, making it highly sensitive to rotational constraints and interparticle interactions.

At volume fractions approaching the maximum packing limit ($\phi \approx 0.90$), pronounced shape-dependent behavior emerges. In this regime, rod suspensions require substantially higher strain thresholds relative to the resuspension process. This divergence reflects fundamental differences in the kinematics of particle motion under strong confinement. While spherical particles can rearrange through relatively simple translational displacements, elongated particles require rotational motion to disengage from the surrounding structure[23,27,34,41]. However, at high $\phi$, the reduced free volume and stronger confinement near maximum packing are expected to favor flow alignment while restricting rotational mobility. Such confinement may reduce the available free volume for reorientation and hinder the collective rearrangements required for mobilization. As a result, larger accumulated strains are required primarily for the subsequent transition toward the fully suspended state.

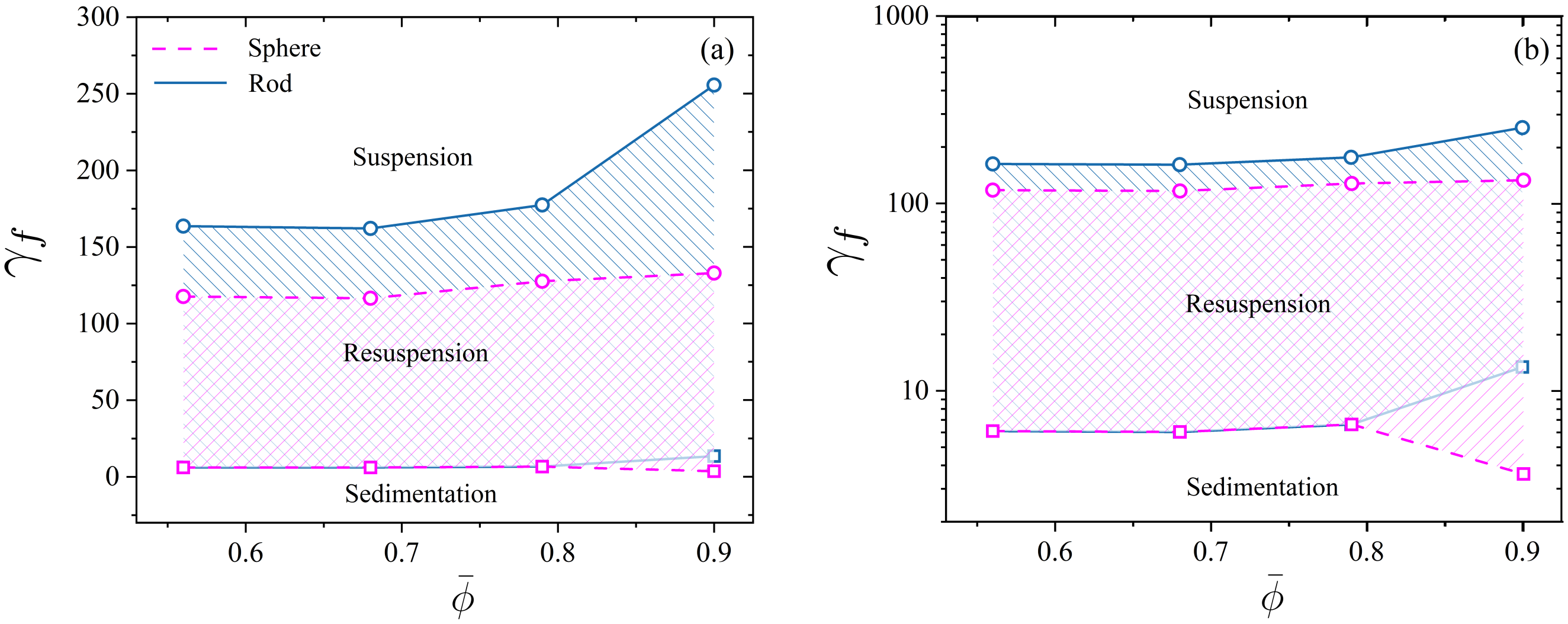

FIG. 10. Fluid strain amplitude, $\gamma_f$, as a function of normalized volume fraction, shown on (a) linear and (b) semi-logarithmic scales. Open squares and circles denote the critical strain thresholds for resuspension initiation, $\gamma_{f,c1}$, and resuspension completion, $\gamma_{f,c2}$, respectively. The shaded regions indicate the strain window over which partial resuspension occurs, with pink and blue corresponding to spherical and rod-shaped particles, respectively.

Although the present study focuses on particles with a single aspect ratio ($\Gamma = 4.6$), the observed increase in $\gamma_{f,c2}$ suggests that the strain required to reach a fully suspended state depends systematically on particle anisotropy. This trend points toward a possible anisotropy-dependent scaling for resuspension, motivating future studies across a broader range of aspect ratios and particle geometries to assess the generality of the strain-controlled framework established here.

While the present orientation analysis provides direct evidence of preferential alignment during resuspension, it represents a two-dimensional projection of the evolving particle structure and does not fully resolve the coupled rotational and translational dynamics within the suspension. Future studies combining time-resolved particle tracking, three-dimensional imaging, or volumetric microscopy with rheological measurements could establish quantitative relationships between orientational order, contact-network evolution, and the strain-controlled transitions identified here. Such measurements would further clarify how particle anisotropy influences the pathway toward a fully suspended state while leaving the onset of resuspension largely unchanged.

## V. CONCLUSION

In this work, we investigated the resuspension behavior of dense non-Brownian suspensions within a strain-controlled rheological framework, with particular emphasis on the role of particle shape. By directly comparing spherical and rod-shaped particles composed of identical materials, the influence of particle morphology was isolated from material-dependent effects. Normalization of the volume fraction using the maximum packing fraction, determined via the Krieger–Dougherty relation, ensured comparable initial sediment structures and enabled meaningful comparisons across particle shapes.

The results demonstrate that viscous resuspension consists of two distinct rheological transitions governed by fluid strain. The onset of resuspension is controlled by a critical strain, with both spherical and rod-shaped suspensions exhibiting nearly identical thresholds over a wide range of concentrations. Particle detachment occurs at a critical fluid strain of approximately $\gamma_{f,c1} \approx 6$, indicating that the initiation of particle motion is governed primarily by a local deformation criterion balancing hydrodynamic forcing and gravitational confinement, and is largely independent of particle morphology.

In contrast, the transition to a fully suspended state depends strongly on particle shape. Spherical suspensions reach a steady state, fully dispersed regime at $\gamma_{f,c2} \approx 120$, whereas rod-shaped suspensions require significantly larger accumulated strain, $\gamma_{f,c2} \approx 180$, with the disparity increasing at high normalized volume fractions. This behavior arises from confinement-induced alignment, restricted rotational mobility, and the additional deformation required for anisotropic particles to undergo microstructural reorganization.

These findings establish a clear mechanistic distinction between particle detachment and complete resuspension: while the former is governed by a nearly consistent strain criterion, the latter is controlled by microstructural evolution and is strongly influenced by particle anisotropy. The additional strain required for rod suspensions is interpreted as reflecting the coupled translational, rotational, and steric rearrangements needed to establish a fully suspended state.

More broadly, these results identify accumulated fluid strain as the governing parameter organizing viscous resuspension across particle morphologies. Particle anisotropy does not fundamentally alter the onset of particle mobilization but instead governs the subsequent structural evolution required to achieve complete suspension. This distinction provides a general framework for understanding resuspension in dense suspensions containing anisotropic particles.

## VI. SUPPLEMENTARY MATERIAL

The supplementary material includes additional figures and detailed calculations that further support the results presented in this study.

## ACKNOWLEDGMENTS

This work was partially supported by USDA award number 2023-67021-39606. P.M. and M.M. acknowledge support in part from the National Science Foundation under Grant No. NSF-CBET-2335195. The authors thank Emilio Burgos for helpful discussions and preliminary image analysis of particle orientation during the early stages of this work.

## DATA AVAILABILITY STATEMENT

The data that support the findings of this study are available from the corresponding author upon reasonable request. All relevant data generated or analyzed during this study are included in this published article.

## REFERENCES

[1] Andrea W Chow, Steven W Sinton, Joseph H Iwamiya, and Thomas S Stephens. Shear-induced particle migration in couette and parallel-plate viscometers: Nmr imaging and stress measurements. *Physics of Fluids*, 6(8):2561–2576, 1994.

[2] JR Abbott, N Tetlow, AL Graham, SA Altobelli, Eiichi Fukushima, LA Mondy, and TS Stephens. Experimental observations of particle migration in concentrated suspensions: Couette flow. *Journal of rheology*, 35(5):773–795, 1991.

[3] Mohammad Sarabian, Mohammadhossein Firouznia, Bloen Metzger, and Sarah Hormozi. Fully developed and transient concentration profiles of particulate suspensions sheared in a cylindrical couette cell. *Journal of Fluid Mechanics*, 862:659–671, 2019.

[4] Andrea W Chow, JH Iwayima, Steven W Sinton, and DT Leighton. Particle migration of non-brownian, concentrated suspensions in a truncated cone-and-plate. In *Society of Rheology Meeting, Sacramento, CA*, volume 103, page 22, 1995.

[5] Eric Brown and Heinrich M Jaeger. The role of dilation and confining stresses in shear thickening of dense suspensions. *Journal of Rheology*, 56(4):875–923, 2012.

[6] FRANCIS ARTURO GADALA-MARIA. *THE RHEOLOGY OF CONCENTRATED SUSPENSIONS.* Stanford University, 1979.

[7] David Leighton and Andreas Acrivos. Viscous resuspension. *Chemical engineering science*, 41(6):1377–1384, 1986.

[8] Andreas Acrivos, Roberto Mauri, and X Fan. Shear-induced resuspension in a couette device. *International journal of multiphase flow*, 19(5):797–802, 1993.

[9] Albert Shields. Anwendung der aehnlichkeitsmechanik und der turbulenzforschung auf die geschiebebewegung. *PhD Thesis Technical University Berlin*, 1936.

[10] Ralph Alger Bagnold. *The physics of blown sand and desert dunes*. Courier Corporation, 2012.

[11] François Charru, Hélene Mouilleron, and Olivier Eiff. Erosion and deposition of particles on a bed sheared by a viscous flow. *Journal of Fluid Mechanics*, 519:55–80, 2004.

[12] Morgane Houssais and E Lajeunesse. Bedload transport of a bimodal sediment bed. *Journal of Geophysical Research: Earth Surface*, 117(F4), 2012.

[13] Morgane Houssais, Carlos P Ortiz, Douglas J Durian, and Douglas J Jerolmack. Onset of sediment transport is a continuous transition driven by fluid shear and granular creep. *Nature communications*, 6(1):6527, 2015.

[14] Marie Lenoble, Patrick Snabre, and Bernard Pouligny. The flow of a very concentrated slurry in a parallel-plate device: Influence of gravity. *Physics of Fluids*, 17(7), 2005.

[15] Malika Ouriemi, Pascale Aussillous, and Elisabeth Guazzelli. Sediment dynamics. part 1. bed-load transport by laminar shearing flows. *Journal of Fluid Mechanics*, 636:295–319, 2009.

[16] Kamyar Najmi, Brenton S McLaury, Siamack A Shirazi, and Selen Cremaschi. The effect of viscosity on low concentration particle transport in single-phase (liquid) horizontal pipes. *Journal of Energy Resources Technology*, 138(3):032902, 2016.

[17] Aman G Kidanemariam and Markus Uhlmann. Interface-resolved direct numerical simulation of the erosion of a sediment bed sheared by laminar channel flow. *International Journal of Multiphase Flow*, 67:174–188, 2014.

[18] Anubhav Tripathi and Andreas Acrivos. Viscous resuspension in a bidensity suspension. *International journal of multiphase flow*, 25(1):1–14, 1999.

[19] Anat Shauly, Amir Wachs, and Avinoam Nir. Shear-induced particle resuspension in settling polydisperse concentrated suspension. *International journal of multiphase flow*, 26(1):1–15, 2000.

[20] G Pp Krishnan and DT Leighton Jr. Dynamic viscous resuspension of bidisperse suspensions—i. effective diffusivity. *International journal of multiphase flow*, 21(5):721–732, 1995.

[21] Mohammadreza Mahmoudian, Simon A Rogers, and Parisa Mirbod. From sedimentation to suspension: Critical strain as a predictor of particle resuspension thresholds. *Journal of Rheology*, 70(4):669–681, 2026.

[22] Laurent Corté, Sharon J Gerbode, Weining Man, and David J Pine. Self-organized criticality in sheared suspensions. *Physical review letters*, 103(24):248301, 2009.

[23] Sebastian Mueller, EW Llewellin, and HM Mader. The rheology of suspensions of solid particles. *Proceedings of the Royal Society A: Mathematical, Physical and Engineering Sciences*, 466(2116):1201–1228, 2010.

[24] S Mueller, EW Llewellin, and HM Mader. The effect of particle shape on suspension viscosity and implications for magmatic flows. *Geophysical Research Letters*, 38(13), 2011.

[25] Nicole M James, Huayue Xue, Medha Goyal, and Heinrich M Jaeger. Controlling shear jamming in dense suspensions via the particle aspect ratio. *Soft matter*, 15(18):3649–3654, 2019.

[26] Eric Brown, Hanjun Zhang, Nicole A Forman, Benjamin W Maynor, Douglas E Betts, Joseph M DeSimone, and Heinrich M Jaeger. Shear thickening and jamming in densely packed suspensions of different particle shapes. *Physical Review E—Statistical, Nonlinear, and Soft Matter Physics*, 84(3):031408, 2011.

[27] M Mahmoudian, F Goharpey, M Behzadnasab, and Z Daneshfar. Shear thickening and hysteresis in dense suspensions: The effect of particle shape. *Journal of Rheology*, 68(3):479–490, 2024.

[28] Ramandeep Jain, Silvio Tschisgale, and Jochen Fröhlich. Impact of shape: Dns of sediment transport with non-spherical particles. *Journal of Fluid Mechanics*, 916:A38, 2021.

[29] E d'Ambrosio, D Gilbert, F Blanc, C Cohen, and Elisabeth Lemaire. Rheology of suspensions of cubic particles. *Journal of Rheology*, 69(6):807–828, 2025.

[30] Leonard H Switzer III and Daniel J Klingenberg. Rheology of sheared flexible fiber suspensions via fiber-level simulations. *Journal of Rheology*, 47(3):759–778, 2003.

[31] Jérôme J Crassous, Lucia Casal-Dujat, Martin Medebach, Marc Obiols-Rabasa, Romaric Vincent, Frank Reinhold, Volodymyr Boyko, Immanuel Willerich, Andreas Menzel, Christian Moitzi, et al. Structure and dynamics of soft repulsive colloidal suspensions in the vicinity of the glass transition. *Langmuir*, 29(33):10346–10359, 2013.

[32] Irvin M Krieger and Thomas J Dougherty. A mechanism for non-newtonian flow in suspensions of rigid spheres. *Trans. Soc. Rheol*, 3(1):137–152, 1959.

[33] Matthieu Wyart and Micheal E Cates. Discontinuous shear thickening without inertia in dense non-brownian suspensions. *Physical review letters*, 112(9):098302, 2014.

[34] Willi Pabst, Eva Gregorová, and Christoph Berthold. Particle shape and suspension rheology of short-fiber systems. *Journal of the European Ceramic Society*, 26(1-2):149–160, 2006.

[35] Subhransu Dhar, Sebanti Chattopadhyay, and Sayantan Majumdar. Signature of jamming under steady shear in dense particulate suspensions. *Journal of Physics: Condensed Matter*, 32(12):124002, 2020.

[36] Bernhard Vowinckel, Edward Biegert, Eckart Meiburg, Pascale Aussillous, and Élisabeth Guazzelli. Rheology of mobile sediment beds sheared by viscous, pressure-driven flows. *Journal of Fluid Mechanics*, 921:A20, 2021.

[37] Aaron PR Eberle, Donald G Baird, Peter Wapperom, and Gregorio M Vélez-García. Obtaining reliable transient rheological data on concentrated short fiber suspensions using a rotational rheometer. *Journal of Rheology*, 53(5):1049–1068, 2009.

[38] David J Pine, Jerry P Gollub, John F Brady, and Alexander M Leshansky. Chaos and threshold for irreversibility in sheared suspensions. *Nature*, 438(7070):997–1000, 2005.

[39] Jonathan M Bricker and Jason E Butler. Oscillatory shear of suspensions of noncolloidal particles. *Journal of rheology*, 50(5):711–728, 2006.

[40] Jonathan M Bricker and Jason E Butler. Correlation between stresses and microstructure in concentrated suspensions of non-brownian spheres subject to unsteady shear flows. *Journal of rheology*, 51(4):735–759, 2007.

[41] Eric Brown, Hanjun Zhang, Nicole A Forman, Benjamin W Maynor, Douglas E Betts, Joseph M DeSimone, and Heinrich M Jaeger. Shear thickening in densely packed suspensions of spheres and rods confined to few layers. *Journal of Rheology*, 54(5):1023–1046, 2010.

[42] George Barker Jeffery. The motion of ellipsoidal particles immersed in a viscous fluid. *Proceedings of the Royal Society of London. Series A, Containing papers of a mathematical and physical character*, 102(715):161–179, 1922.

[43] S. G. Advani and C. L. III Tucker. The use of tensors to describe and predict fiber orientation in short fiber composites. *Journal of Rheology*, 31:751–784, 1987.

[44] F. Folgar and C. L. Tucker. Orientation behavior of fibers in concentrated suspensions. *Journal of Reinforced Plastics and Composites*, 3:98–119, 1984.

[45] F. P. Bretherton. The motion of rigid particles in a shear flow at low reynolds number. *Journal of Fluid Mechanics*, 14:284–304, 1962.